\documentclass[twocolumn,pre,showpacs,amsfonts,amsmath,assymb,eufrak]{revtex4-1}
\usepackage{epsfig}
\usepackage{color}
\usepackage{bm}
\usepackage{braket}
\usepackage{graphicx}
\usepackage{amsthm}

\usepackage{calc}

\newcommand{\eq}[1]{\begin{equation} #1 \end{equation}}
\newcommand{\eqa}[2]{\begin{equation} #1 \label{#2} \end{equation}}
\newcommand{\balign}[1]{\begin{align} #1 \end{align}}

\newcommand{\figin}[4]
{\begin{figure}[tb]
\centering
\includegraphics[width= #1]{#2.pdf}
\caption{#3}
\label{f:#4}
\end{figure}}

\newcommand{\todayd}{\the\year/\the\month/\the\day}

\newcommand{\bib}{\bibitem}

\newcommand{\lb}{\label}

\newcommand{\bel}{\begin{easylist}}
\newcommand{\eel}{\end{easylist}}

\newcommand{\be}[1]{\begin{enumerate} #1 \end{enumerate}}

\newcommand{\eref}[1]{Eq.~\eqref{#1}}

\newcommand{\fref}[1]{Fig.~\ref{f:#1}}

\def \({\left(}
\def \){\right)}
\def \[{\left[}
\def \]{\right]}
\newcommand{\la}{\langle}
\newcommand{\ra}{\rangle}

\newcommand{\sumtwo}[2]%
{\mathop{\sum_{#1}}_{#2}}
\newcommand{\sumthree}[3]%
{\mathop{\mathop{\sum_{#1}}_{#2}}_{#3}}
\newcommand{\sumfour}[4]%
{\mathop{\mathop{\mathop{\sum_{#1}}_{#2}}_{#3}}_{#4}} 
\newcommand{\prodtwo}[2]%
{\mathop{\prod_{#1}}_{#2}}
\newcommand{\mintwo}[2]%
{\mathop{\min_{#1}}_{#2}}
\newcommand{\maxtwo}[2]%
{\mathop{\max_{#1}}_{#2}}
\newcommand{\maxthree}[3]%
{\mathop{\mathop{\max_{#1}}_{#2}}_{#3}}
\newcommand{\limtwo}[2]%
{\mathop{\lim_{#1}}_{#2}}
\newcommand{\suptwo}[2]%
{\mathop{\sup_{#1}}_{#2}}
\newcommand{\supthree}[3]%
{\mathop{\mathop{\sup_{#1}}_{#2}}_{#3}}
\newcommand{\supfour}[4]%
{\mathop{\mathop{\mathop{\sup_{#1}}_{#2}}_{#3}}_{#4}} 
\newcommand{\inftwo}[2]%
{\mathop{\inf_{#1}}_{#2}}
\newcommand{\infthree}[3]%
{\mathop{\mathop{\inf_{#1}}_{#2}}_{#3}}
\newcommand{\inffour}[4]%
{\mathop{\mathop{\mathop{\inf_{#1}}_{#2}}_{#3}}_{#4}} 

\newcommand\calZ{{\cal Z}}

\newcommand{\Di}{\mathit{\Delta}}

\newcommand{\para}[1]{{\em #1}\/.---}

\def\rnum#1{\resizebox{0.5em}{\height}{\expandafter{\romannumeral #1}}}
\def\Rnum#1{\resizebox{0.5em}{\height}{\uppercase\expandafter{\romannumeral #1}}}

\newcommand{\JA}{J_{\rm A}}
\newcommand{\JB}{J_{\rm B}}
\newcommand{\IA}{I_{\rm A}}
\newcommand{\IB}{I_{\rm B}}
\newcommand{\DA}{d_{\rm A}}
\newcommand{\DB}{d_{\rm B}}
\newcommand{\dsgm}{\dot{\sigma}}
\newcommand{\Var}{{\rm Var}}

\newcommand{\rac}{\rangle_{\rm c}}
\newcommand{\DmuA}{\Di \mu_{\rm A}}
\newcommand{\DmuB}{\Di \mu_{\rm B}}

\newcommand{\ChiA}{\chi_{\rm A}}
\newcommand{\ChiB}{\chi_{\rm B}}

\makeatletter
\renewcommand{\@cite}[1]{\textsuperscript{#1)}}
\makeatother

\begin{document}

\preprint{APS/123-QED}

\newcommand{\titlehere}{Is there any upper bound on current fluctuation in cross-transport beyond linear response regime?}

\newcommand{\titlename}{\titlehere}

\preprint{APS/123-QED}

\title{\titlename}

\author{Naoto Shiraishi$^{1}$, Tan Van Vu$^{2}$, and Keiji Saito$^{2}$ }
\affiliation{$^{1}$Department of Physics, Gakushuin University, 1-5-1 Mejiro, Toshima-ku, Tokyo 171-8588, Japan} 

\affiliation{$^{2}$Department of Physics, Keio University, 3-14-1 Hiyoshi, Kohoku-ku, Yokohama 223-0061, Japan}%

\date{\today}

\begin{abstract}
Input and output current fluctuations in stationary cross-transport systems with two kinds of currents are investigated.
In Saryal {\it et al.} [Phys. Rev. Lett. 127, 190603 (2021)], upper bounds on input and output current fluctuations are proven in the linear response regime and suggested the validity beyond the linear response regime.
We examine these bounds beyond the linear response regime and find that these bounds are violated in general nonequilibrium stationary conditions.
We show two examples to demonstrate the violation. 
Through these examples, we argue that no upper bound on input and output current fluctuations exist beyond the linear response regime.
\end{abstract}

\maketitle


\section{Introduction}

Fluctuation is one of the most important quantities in nonequilibrium statistical mechanics.
A milestone is the fluctuation-response relation in the linear response regime~\cite{Joh28}, which connects the fluctuation of currents to the response of currents to small external perturbations.
The fluctuation-response relation was first proved on the basis of the consistency with the second law of thermodynamics~\cite{Nyq28}, and later its microscopic foundation was revealed~\cite{Tak52, Kub57}.
The fluctuation theorem~\cite{ECM93, GC95, Jar97, Kur98} reproduces the fluctuation-response relation as its low-order description~\cite{ES95, Gal96}.
Applying this technique, higher-order extensions and further generalizations of fluctuation-response relation have been obtained~\cite{SD,SU,Nak11, APE16, Shi22}.

These equalities are obtained around zero current, and the fluctuation of current around a finite amount of current is not easy to characterize.
Recently-obtained relations on current fluctuations thus take the form of inequalities, not equalities, to evaluate the fluctuation of current in general stationary systems.
The thermodynamic uncertainty relation is a prominent example of this research direction, which provides a lower bound of the relative fluctuation (i.e., fluctuation divided by the square of its average) of an arbitrary current by the inverse of entropy production~\cite{BS15, Ging16, GRH17, DS20}.
Although this inequality does not achieve its equality in general nonequilibrium setups~\cite{Shi21}, the thermodynamic uncertainty relation is considered to be a good clue to understand nonequilibrium stationary systems~\cite{HG20}, and various extensions~\cite{BHS18, DS18, VH19, HV19, LGU20, KS20, VVH20, DS21, VS21, Hase21, LPP21} and similar inequalities connecting fluctuation and entropy production~\cite{SST16, PS18, SFS18, SS19} have been investigated intensively.

The aforementioned results uncover several {\it lower} bounds on the fluctuation of currents.
Very recently, two {\it upper} bounds on the fluctuation of currents were proposed.
These inequalities apply to stationary cross-transport systems where current A flows along thermodynamic force and current B flows against thermodynamic force.
Currents A and B can be regarded as a fuel (free energy consumption) and a load (work extraction), respectively.
The first inequality claims that the fluctuation of current B is smaller than A, and the second inequality claims that the relative fluctuation of current A is smaller than B.
The second inequality was first proposed by Ito {\it et al.}~\cite{Ito19} in a cyclic heat engine in a slightly different form, and then Saryal {\it et al.}~\cite{Sar21} formulated these two inequalities for general setups and proved them in the linear response regime with time-reversal symmetry.
They also demonstrated the validity of these inequalities beyond the linear response regime by numerical simulations.
These inequalities are also verified in cyclic heat engines~\cite{Ito19, Sar21b} and generalized stationary setups in the linear response regime~\cite{Sar21c, Moh21}.

Motivated by these observations, in this paper, we investigate an upper bound on currents in stationary cross-transport systems beyond the linear response regime.
We consider two simple models to revisit the inequalities proposed in \cite{Sar21}.
  The first model is a chemical reaction system consisting of several reactions with two types of particles, and the second model is a thermoelectric transport system.
 In the first model, we examine the stationary state and find that the inequalities are violated in a general nonequilibrium condition beyond the linear response regime, which is shown by constructing concrete
setups.
  A key observation behind this construction is the existence of a reaction with negligible current but infinitely large current fluctuation.
Through this model, we
  establish that the fluctuation of one of the currents in the cross transport cannot be bounded from above by average currents and the fluctuation of another current. 
Moreover, although the above construction in the first model requires highly nonequilibrium conditions, we numerically demonstrate that these inequalities are also frequently violated in mild nonequilibrium conditions by using the second model, i.e., the thermoelectric transport system.
This numerical simulation clearly shows that these inequalities are violated in a wide parameter range, including the case that the corresponding heat engine is not inefficient.

\section{Input-output fluctuation inequalities}

We first explain the input-output fluctuation conjecture, which was given by Saryal {\it et al.}~\cite{Sar21}.
They considered a stationary system with two currents $\JA$ and $\JB$, which transport two different matters, A and B.
For instance, A is heat and B is electrons. Another example is that A and B are two different kinds of particles.
For the convenience of explanation, we shall employ the description with the latter setup, while our results directly apply to the former cases.

We denote the difference of their conjugate intensive variable by $\DA$ and $\DB$, respectively.
The contributions from current A and B to entropy production rate are given by $\IA:=\DA\JA$ and $\IB:=\DB\JB$, whose sum is equal to the total entropy production rate $\dsgm=\IA+\IB$.

Suppose that A is fuel, $\IA>0$, and B is work extraction, $\IB<0$.
We call these two currents $\IA$ and $\IB$ also as the input current and the output current, respectively.
We express the fluctuation (the scaled second cumulant) of a current as $\la \Di I^2\rac=\lim_{\tau\to \infty}  \tau \left\la \(\frac1\tau \int_0^\tau dt I(t)-\la I\ra \)^2\right\ra $.
Saryal {\it et al.}~\cite{Sar21} claimed the following two inequalities on the fluctuations of input and output currents.
\be{
\item The fluctuation of the output current $\IB$ should be smaller than that of the input current $\IA$:
\eqa{
\la \Di \IB^2 \rac \leq \la \Di \IA^2 \rac . 
}{conjecture1}
\item The relative fluctuation of the output current $\IB$ should be larger than that of the input current $\IA$:
\eqa{
 \frac{\la \Di \IA^2\rac}{\IA^2}\leq \frac{\la \Di \IB^2\rac}{\IB^2}.
}{conjecture2}
}
Saryal {\it et al.} proved these inequalities in systems with time-reversal symmetry in the linear response regime.
In addition, on the basis of numerical simulations, they strongly suggest the validity of these inequalities in general nonequilibrium stationary systems.
Note that their proof in the linear response regime is based on the linear expansion of currents with the Onsager matrix and thus cannot  be extended to a nonlinear regime directly.

\section{First counterexample: Chemical reactions with Poisson processes}

Contrary to the aforementioned expectation, we shall construct concrete counterexamples to these two inequalities beyond the linear response regime.
We also show that there exists no upper bound on a current fluctuation as long as we use the average currents and another current fluctuation.

\subsection{Setup}

We consider a transport system with a single internal state attached to four particle baths; two for particle A and the other two for particle B.
Particles are carried by reactions that occur inside the system.
The whole system is in an isothermal condition, and reactions are driven by consuming chemical potential.
As an example, in reaction 1 three particles A are transported from bath HA to bath LA, and at the same time, four particles B are transported from bath LB to HB (see \fref{schematic}).

Remark that this formulation is not restricted to chemical reactions but also general Markov processes with a single internal state.
Another example is a multiterminal ballistic transport in the semiclassical regime~\cite{BHS18}.

\figin{8.5cm}{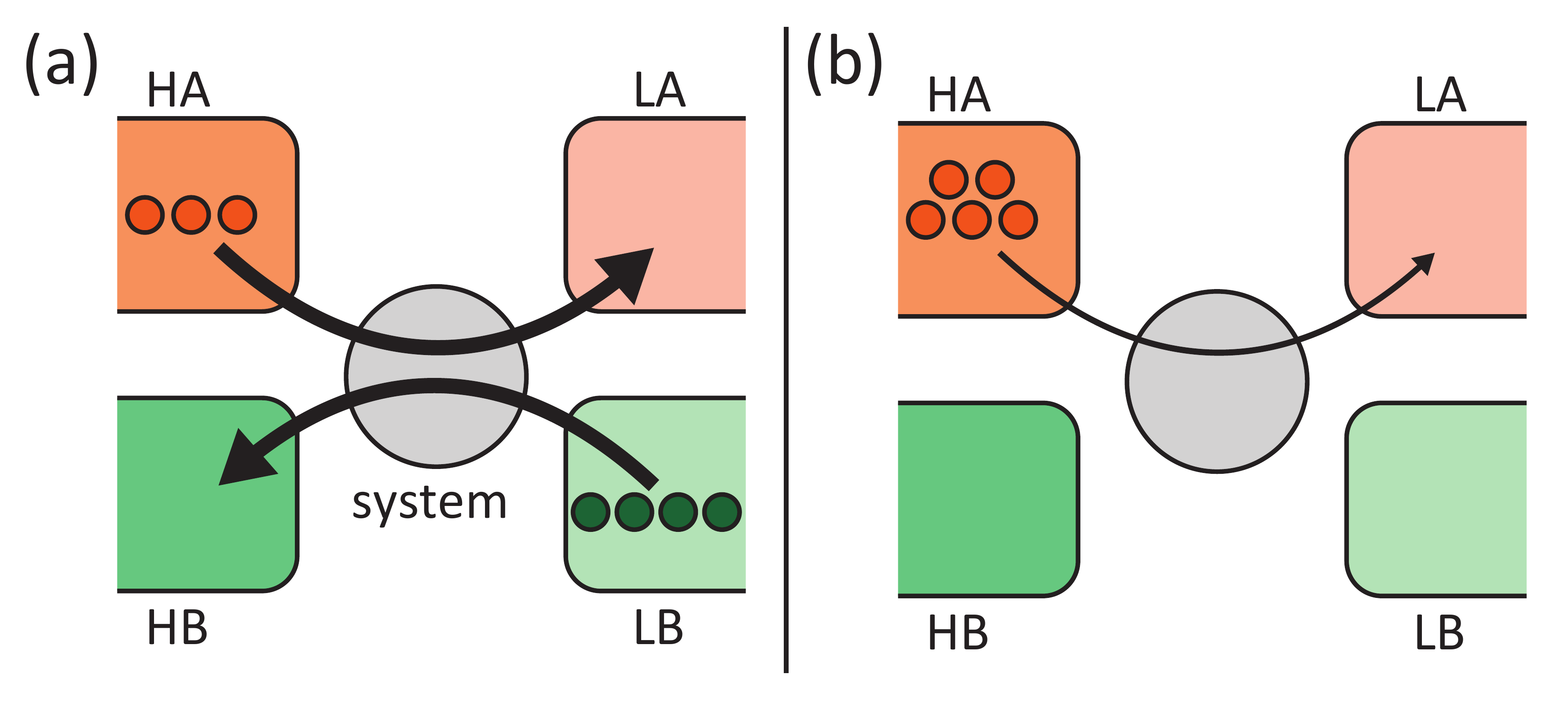}{
(a) An example of a reaction, referred to as reaction 1 in the main text.
In this reaction, three particles A flow from particle bath HA to HL, which induces transport of four particles B from LB to HB.
(b) A schematic of the key reaction to construct counterexamples.
This reaction concerns only one of particles A or B.
This reaction rarely occurs, while it transports many particles.
}{schematic}

Let $a_i$ and $b_i$ be the entropy production associated with particles A and B in the $i$-th reaction, respectively.
In our setup, these quantities are given by $a_i=\beta \Di \mu_{\rm A} n_i^{\rm A}$ and $b_i=\beta \Di \mu_{\rm B} n_i^{\rm B}$, where $n_i^{\rm A}$ (resp. $n_i^{\rm B}$) is the number of particles A (resp. B) transported from HA (resp. HB) to LA (resp. LB) in the $i$-th reaction, and $\Di \mu_{\rm A}:=\mu_{\rm HA}-\mu_{\rm LA}$ (resp. $\Di \mu_{\rm B}:=\mu_{\rm HB}-\mu_{\rm LB}$) is the chemical potential difference between HA and LA (resp. HB and LB).
In this reaction system, the combination $(n_i^A , n_i^B)$ is fixed for each reaction $i$.
If particles flow from LA to HA (resp. LB to HB) in reaction $i$, we set negative $n_i^{\rm A}$ (resp. $n_i^{\rm B}$).
We assume that all reactions follow the Poisson processes.
Due to the local detailed-balance condition, the jump rate of the $i$-th forward and backward reactions are written as
\balign{
P_i^{\rm F}&=T_i e^{a_i+b_i}, \\
P_i^{\rm B}&=T_i,
}
with $T_i\geq 0$. 
Here, the forward and backward reactions accompany the same transition coefficient $T_i$ due to the time-reversal symmetry of our system.
The entropy production rate associated with particle A is written as
\eq{
\IA=\sum_i a_i (P_i^{\rm F}-P_i^{\rm B})=\sum_i T_i a_i(e^{a_i+b_i}-1).
}
With noting $\IA=\beta \DmuA \JA$, the above expression in the linear response regime $\Di \mu_{\rm A}, \Di \mu_{\rm B}\to 0$ suggests
\eq{
\JA= \sum_i  \beta T_i [(n_i^{\rm A})^2\DmuA+n_i^{\rm A}n_i^{\rm B}\DmuB],
}
which defines the Onsager matrix.

The second cumulants of the $i$-th forward and backward processes are also given by $T_i e^{a_i+b_i}$ and $T_i$, which follows from a basic property of a Poisson process that all the cumulants of a Poisson process are the same.
With recalling that the variance of the sum of two independent stochastic variables is equal to the sum of the variances of these two;
\eq{
\Var (X\pm Y)=\Var (X)+\Var (Y),
}
the fluctuation of $\IA$ and $\IB$ (i.e., scaled variance by time $\tau$) is calculated as
\balign{
\la \Di \IA^2\rac&=\sum_i T_i a^2_i(e^{a_i+b_i}+1), \\
\la \Di \IB^2\rac&=\sum_i T_i b^2_i(e^{a_i+b_i}+1).
}
Then, the inequalities \eqref{conjecture1} and \eqref{conjecture2} in this reaction system read
\balign{
\sum_i T_i b^2_i(e^{a_i+b_i}+1)&\leq \sum_i T_i a^2_i(e^{a_i+b_i}+1), \lb{con1} \\
\frac{\sum_i T_i a^2_i(e^{a_i+b_i}+1)}{\( \sum_i T_i a_i(e^{a_i+b_i}-1)\) ^2}&\leq \frac{\sum_i T_i b^2_i(e^{a_i+b_i}+1)}{\( \sum_i T_i b_i(e^{a_i+b_i}-1)\) ^2}, \lb{con-2}
}
under the conditions $\sum_i T_i a_i(e^{a_i+b_i}-1)>0$ and $\sum_i T_i b_i(e^{a_i+b_i}-1)<0$. For the sake of notational simplicity, we introduce the following symbols:
\balign{
C_{1,i}^X:=&T_i X_i(e^{a_i+b_i}-1), \\
C_{2,i}^X:=&T_i X^2_i(e^{a_i+b_i}+1),
}
with $X=a,b$. Then, these inequalities are written in a concise form:
\balign{
  \sum_i C_{2,i}^b\leq& \sum_i C_{2,i}^a,  \lb{con1C} \\
\frac{\sum_i C_{2,i}^a}{(\sum_i C_{1,i}^a)^2}\leq& \frac{\sum_i C_{2,i}^b}{(\sum_i C_{1,i}^b)^2} \, , \lb{con2C} 
}
under the conditions $ \sum_i C_{1,i}^a >0 $ and $ \sum_i C_{1,i}^b <0$.

\subsection{Construction of counterexamples}

A crucial observation for the violation of inequalities \eqref{conjecture1} and \eqref{conjecture2} is the following:
There exists a reaction concerning only A (i.e., particle B is completely irrelevant) such that $C_{1,i}^a$ vanishes while $C_{2,i}^a$ diverges in some limit (see \fref{schematic}.(b)).
Obviously, similar reactions exist for particle B.

We set $T_i$ as
\eq{
  T_i=\frac{k}{(a_i)^{3/2}e^{a_i}}
}
with a constant $k$, and take $a_i$ sufficiently large.
Note that by changing $n_i^{\rm A}$ we can prepare arbitrarily large $a_i$ with keeping other $a_j$'s.
Taking the $a_i\to \infty$ limit, we have
\balign{
\lim_{a_i\to \infty}C_{1,i}^a&=\lim_{a_i\to \infty}\frac{ k } {\sqrt{a_i}}(1-e^{-a_i})=0, \\
\lim_{a_i\to \infty}C_{2,i}^a&=\lim_{a_i\to \infty} k \sqrt{a_i}(1+e^{-a_i})\to +\infty .
}
We name this reaction a {\it spike reaction}.

A counterexample to \eref{conjecture2} is a reaction system with two reactions:
We set reaction 1 as a normal cross-transport reaction satisfying $C_{1,1}^a, C_{1,1}^b, C_{2,1}^a, C_{2,1}^b=O(1)$ with $C_{1,1}^a>0$ and $C_{1,1}^b<0$, and reaction 2 as the spike reaction.
By setting $a_i$ sufficiently large, the left-hand side of \eref{con2C} can become arbitrarily large with keeping its right-hand side at $O(1)$, which violates \eref{con2C} (i.e., \eref{conjecture2}).

A counterexample to \eref{conjecture1} is a reaction system with two reactions:
We again set reaction 1 as a normal cross-transport reaction satisfying $C_{1,1}^a, C_{1,1}^b, C_{2,1}^a, C_{2,1}^b=O(1)$ with $C_{1,1}^a>0$ and $C_{1,1}^b<0$, and reaction 2 as the spike reaction concerning particle B.
By setting $b_i$ sufficiently large, positive work can be extracted (i.e., $C_{1,1}^b+C_{1,2}^b<0$) and the left-hand side of \eref{con1C} can become arbitrarily large with keeping its right-hand side at $O(1)$, which violates \eref{con1C} (i.e., \eref{conjecture1}).

We verify how large nonequilibrium driving is needed to violate the inequalities.
We set $T_1=1$, $a_1=2$, and $b_1=-1$, which are realized with $\beta\DmuA=\beta\DmuB=1$, $n_1^{\rm A}=2$, and $n_1^{\rm B}=-1$.
We first demonstrate the violation of \eref{conjecture2}.
We set reaction 2 as the spike reaction concerning particle A with $k=1/2$.
Then, \eref{conjecture2} reads
\eq{
\frac{4(e+1)+k\sqrt{a}(1+e^{-a})}{\( 2(e-1)+\frac{k}{\sqrt{a}}(1-e^{-a})\) ^2} \leq \frac{e+1}{(e-1)^2}\simeq 1.259.
}
On the other hand, the left-hand side takes $1.2611\cdots$ with $a=9$ (i.e., $n_2^{\rm A}=9$), which exceeds the right-hand side $1.2593\cdots$.

We next demonstrate the violation of \eref{conjecture1}.
We set reaction 2 as the spike reaction concerning particle B with $k=4.3$.
Then, \eref{conjecture1} reads
\eq{
e+1+k\sqrt{b}(1+e^{-b})\leq 4(e+1)\simeq 14.9.
}
On the other hand, the left-hand side takes $15.10\cdots$ with $b=7$ (i.e., $n_2^{\rm B}=7$), which exceeds the right-hand side $14.87\cdots$, with keeping current B negative: $\IB=-(e-1)+\frac{k}{\sqrt{b}}(1-e^{-b})=-0.0945\cdots$.

\bigskip

We shall clarify the physical picture of the spike reaction, which plays a crucial role in the counterexamples.
The spike reaction is a kind of leakage which occurs very rarely but accompanies extremely large entropy production once it happens.
The balance between these two (small probability and large entropy production) provides negligible average and diverging fluctuation.
Here, if we consider thermoelectric transport instead of cross-transport of two particle currents, the spike reaction is interpreted as a rare stochastic path with a large amount of heat transport.
We remark that our construction requires finite $\DmuA$ (or $\DmuB$) and large $n_i^{\rm A}$ (or $n_i^{\rm B}$), and thus this counterexample works only beyond the linear response regime.

The existence of such a leakage reaction suggests that we cannot expect modified versions of inequalities such as $\la \Di \IA\rac \geq \eta \la \Di \IB\rac$ and $\la \Di \IB^2\rac/\IB^2\geq \eta\cdot{\la \Di \IA^2\rac}/{\IA^2}$ with the efficiency $\eta:=-\IB/\IA$.
More generally, we claim that there is no upper bound of  $\la \Di \IA^2\rac$ and  $\la \Di \IB^2\rac$ in the following form:
\balign{
\la \Di \IA^2\rac &\leq f(\IA, \IB, \la \Di \IB^2\rac), \\
\la \Di \IB^2\rac &\leq g(\IA, \IB, \la \Di \IA^2\rac),
}
where $f$ and $g$ are functions with no singular points.
This theorem is readily proven in the following manner:
By employing the spike reaction, we have a diverging  $\la \Di \IA^2\rac$ (resp. $\la \Di \IB^2\rac$) with keeping $\IA$, $\IB$, and $\la \Di \IB^2\rac$ (resp. $\la \Di \IA^2\rac$) at $O(1)$, which clearly violates these inequalities.

\section{Second example: Overdamped thermoelectric transport}

Although we have constructed explicit counterexamples to Eqs.~\eqref{conjecture1} and \eqref{conjecture2}, one may still feel that the violation of Eqs.~\eqref{conjecture1} and \eqref{conjecture2} is very rare phenomena.
To examine how frequently or how rarely these inequalities are violated,
we consider a simple thermoelectric device introduced in Ref.~\cite{Rutten07}, which can be described as an overdamped probabilistic process.
This device transports electrons between two leads through a two-level quantum dot.
We assume that there is always at most one electron in the quantum dot due to the repulsion between electrons.
Consequently, the quantum dot can take three possible states: it is either empty (state $0$) or contain one electron in energy level $E_1$ (state $1$) or $E_2~(>E_1)$ (state $2$).
Each energy level $E_i$ is connected to lead $i$ with chemical potential $\mu_i~(\mu_2>\mu_1)$ and temperature $T_c$.
The transition rates describing the electron exchange between the leads and the quantum dot are given by
\begin{align}
w_{i0} &= \gamma_if(x_i), \\
w_{0i} &= \gamma_i[1 - f(x_i)],
\end{align}
where $\gamma_i>0$ denotes the coupling strength to lead $i$, $f(x)=1/(1+e^x)$ is the Fermi distribution, and we defined $x_i:=(E_i-\mu_i)/T_c$.
The electron transitions between state $1$ and state $2$ are mediated by two heat baths; a cold bath at temperature $T_c$ and a hot bath at temperature $T_h~(>T_c)$.
These transition rates are given by
\begin{align}
w_{12}&=w_{12}^c+w_{12}^h,\\
w_{21}&=w_{21}^c+w_{21}^h,
\end{align}
where we set $w_{21}^a=\gamma_an(x_a)$, $w_{12}^a=\gamma_a[n(x_a)+1]$ with $n(x):=1/(e^x-1)$ as the Bose-Einstein distribution.
We here defined $x_a:=(E_2-E_1)/T_a$ for $a\in\{c,h\}$ and $\gamma_c$ and $\gamma_h$ as the coupling strength to the heat baths.
Notice that the symbols $c$ and $h$ correspond to the cold and hot heat baths.
Operationally, this device can work as a heat engine that converts part of the heat absorbed from the hot heat bath into work in the form of the transport of electrons from lower to higher chemical potentials.

Let $\ket{p_t}=[p_0(t),p_1(t),p_2(t)]^\top$ be the probability distribution of the quantum dot at time $t$, then its time evolution can be described by the master equation
\begin{equation}
\ket{\dot p_t}=\mathsf{W}\ket{p_t},
\end{equation}
where $\mathsf{W}=[w_{ij}]\in\mathbb{R}^{3\times 3}$ satisfies the normalization condition, $\sum_iw_{ij}=0$.
Hereinafter, we exclusively focus on the steady-state dynamics of the system.
The steady-state distribution $\ket{p^{\rm ss}}$ is unique and can be obtained by solving equation $\mathsf{W}\ket{p^{\rm ss}}=0$.

\figin{8.5cm}{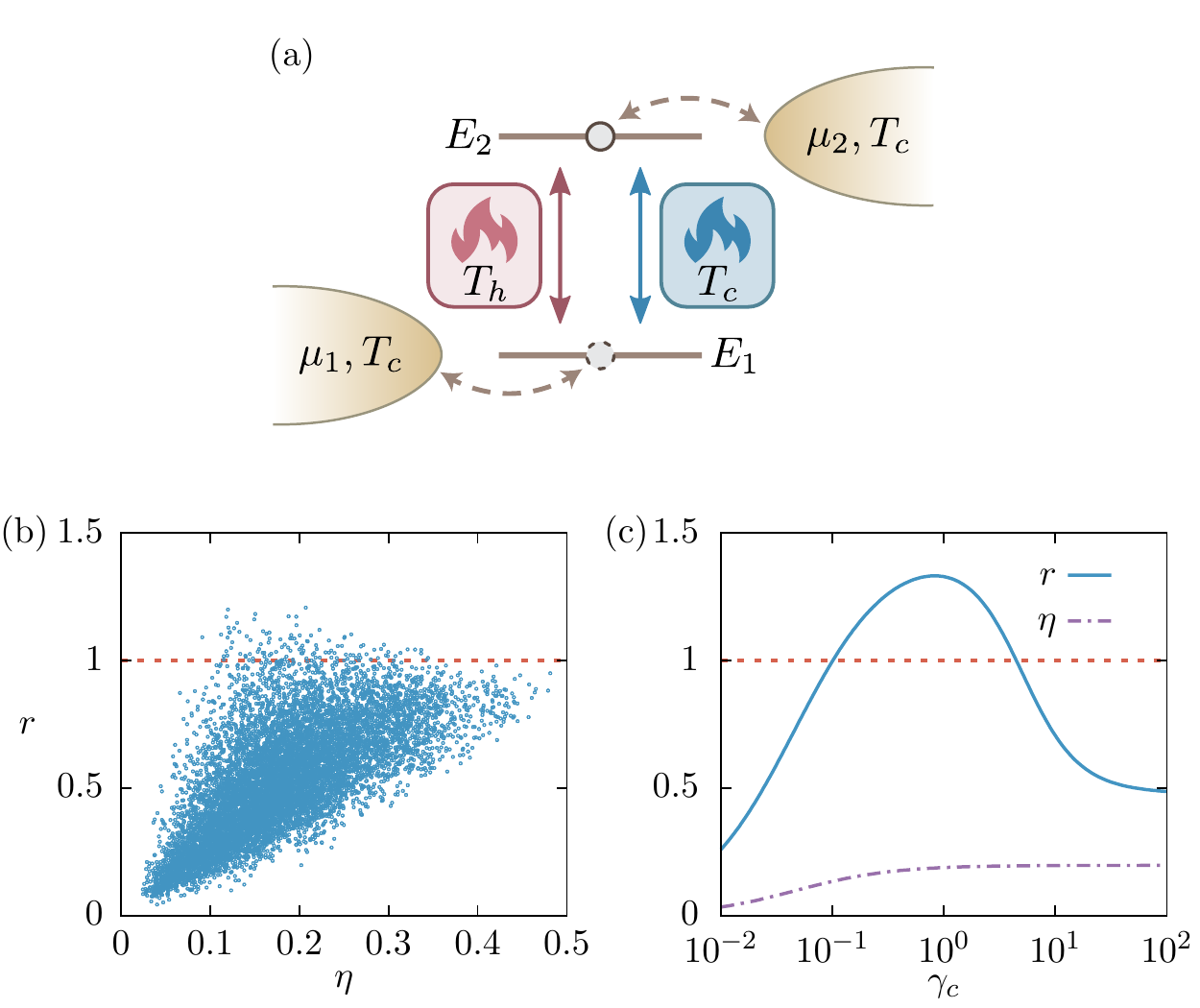}{
(a) A schematic of the thermoelectric device.
Two leads supply electrons to a two-level system.
The transitions between these two levels are driven by two heat baths with different temperatures.
(b) Numerical results of random sampling. 
A single dot depicts a point $(\eta,r)$ for each parameter, and the dashed line depicts the upper bound of $r$ according to Eq.~\eqref{conjecture2}. The parameter ranges are $\beta_h\in[0.1,0.5]$, $\beta_c\in[5,10]$, $\gamma_i,\gamma_a\in[1,10]$, $\mu_1\in[0.1,0.3]$ and $\mu_2\in[0.5,0.7]$. Here, $\beta_h$ and $\beta_c$ are inverse temperatures of $T_h$ and $T_c$, respectively.
(c) Numerical results as changing $\gamma_c$. Plotted are ratio $r$ (solid line) and efficiency $\eta$ (dash-dotted line). The other parameters are fixed as $\beta_h=0.1$, $\beta_c=10$, $\gamma_1=\gamma_2=\gamma_h=10$, $\mu_1=0.1$ and $\mu_2=0.5$.
}{thermoelectric}

To examine Eq.~\eqref{conjecture2}, we consider the input and output currents as follows:
\begin{align}
\JA&=(E_2-E_1)( w_{21}^hp_1^{\rm ss} - w_{12}^hp_2^{\rm ss} ),\\
\JB&=w_{10}p_0^{\rm ss} - w_{01}p_1^{\rm ss}.
\end{align}
Specifically, $\JA$ is the heat flux from the hot heat bath to the system and $\JB$ is the electron flux from lead $1$ to lead $2$.
The total entropy production rate reads \cite{Rutten07}
\begin{equation}
\dot\sigma=\JA \DA+\JB \DB=\IA + \IB,
\end{equation}
where $\DA=1/T_c-1/T_h$ and $\DB=(\mu_1-\mu_2)/T_c$ are the thermodynamic forces conjugated to $\JA$ and $\JB$, respectively.
As long as the device works as a heat engine, we always have $\IA>0$ and $\IB<0$.

The finite-time fluctuations of $\IA$ and $\IB$ can be calculated by means of full counting statistics.
Introducing the counting fields $\ChiA$ and $\ChiB$, the characteristic function is written as
\begin{equation}
\calZ(\ChiA,\ChiB)=\bra{1}e^{\mathsf{W}(\ChiA,\ChiB)\tau}\ket{p^{\rm ss}},
\end{equation}
where $\bra{1}:=[1,1,1]$ is the all-one vector and the modified rate matrix is given by
\begin{equation}
\mathsf{W}(\ChiA,\ChiB)=\begin{bmatrix}
	w_{00} & w_{01}e^{-\ChiB} & w_{02}\\
	w_{10}e^{\ChiB} & w_{11} & w_{12}^c+w_{12}^he^{-\ChiA}\\
	w_{20} & w_{21}^c+w_{21}^he^{\ChiA} & w_{22}
\end{bmatrix}.
\end{equation}
Using the characteristic function $\calZ(\ChiA,\ChiB)$, the fluctuations of the input and output currents can thus be calculated using $\calZ(\ChiA,\ChiB)$ as follows:
\begin{align}
\la \Di \IA^2\rac &=(\DA\Delta E)^2\left.{\frac{\partial^2}{\partial\ChiA^2}\lim_{\tau\to\infty}\frac{1}{\tau}\ln\calZ(\ChiA,0)}\right|_{\ChiA=0},\label{eq:fluctuation.A}\\
\la \Di \IB^2\rac &=(\DB)^2\left.{\frac{\partial^2}{\partial\ChiB^2}\lim_{\tau\to\infty}\frac{1}{\tau}\ln\calZ(0,\ChiB)}\right|_{\ChiB=0},\label{eq:fluctuation.B}
\end{align}
where $\Delta E=E_2-E_1$.
These quantities can be easily evaluated using the dominant eigenvalue of $\mathsf{W}(\ChiA,\ChiB)$ with the largest real part.

Now we shall numerically test Eq.~\eqref{conjecture2} with changing parameters.
First, we randomly sample the temperatures $T_c$ and $T_h$, coupling strengths $\gamma_i$ and $\gamma_a$, chemical potentials $\mu_i$ ($i=1,2$ and $a\in\{c,h\}$), while the energy levels $E_i$ are fixed as $E_1=0$ and $E_2=1$.
For each parameter setting, the fluctuations of the input and output currents in Eqs.~\eqref{eq:fluctuation.A} and \eqref{eq:fluctuation.B} are calculated using numerical differentiation at $\epsilon=10^{-4}$.
Note that we exclude all parameter samples where the device does not function as a heat engine.
Specifically, only parameters that result in $\IA>0$ and $\IB<0$ are retained.
In Fig.~\ref{f:thermoelectric}(b), we plot all selected points $(\eta,r)$ in a two-dimensional plane, where $\eta$ is the efficiency of the device and $r$ is the ratio defined as
\begin{equation}
r=\frac{\la \Di \IA^2\rac}{\IA^2}\left(\frac{\la \Di \IB^2\rac}{\IB^2}\right)^{-1},
\end{equation}
which should be smaller than or equal to $1$, according to Eq.~\eqref{conjecture2}.
As can be seen, there are many points above the line $r=1$, which implies that Eq.~\eqref{conjecture2} is violated.

Next, in order to make clear when Eq.~\eqref{conjecture2} is violated, we fix all parameters except for $\gamma_c$, which is varied from $10^{-2}$ to $10^{2}$.
In this far-from-equilibrium regime, we confirm that the device always works as a heat engine.
We calculate $r$ and $\eta$ as functions of $\gamma_c$, which is plotted in Fig.~\ref{f:thermoelectric}(c).
It can be verified that Eq.~\eqref{conjecture2} does not hold (i.e., $r>1$) for $\gamma_c\in(0.1,4)$.
In this parameter range, the efficiency $\eta$ of the device is about $0.19$, which is not so inefficient.
This implies that Eq.~\eqref{conjecture2} can be easily violated in a wide range where the device can be operated as a useful heat engine.

\section{Conclusion}

We have shown that two inequalities \eqref{conjecture1} and \eqref{conjecture2} are violated in general nonequilibrium conditions by explicitly constructing counterexamples to these inequalities.
The crucial element of our counterexample is the spike reaction, which has arbitrarily small current and arbitrarily large current fluctuation.
Our counterexamples not only refute these conjectures but also deny the existence of upper bounds on current fluctuation in a certain form.

Our result suggests that one possible direction to obtain upper bounds on current fluctuation is to take higher-order cumulants of currents.
Utilization of higher-order cumulants in stochastic thermodynamics has recently been attempted in the context of thermodynamic uncertainty relations~\cite{DS20, Kam21} and reaction networks~\cite{BS15b}.
With the help of these results, we may reach useful upper bounds of current, which is left as a future problem.

\bigskip

\para{Acknowledgement}
NS is supported by JSPS KAKENHI Grants-in-Aid for Early-Career Scientists Grant Number JP19K14615. KS is supported
by Grants-in-Aid for Scientific Research Grant Number JP19H05603 and JP19H05791.

\end{document}